\documentclass[aps,prl,twocolumn,superscriptaddress,showpacs,floatfix]{revtex4-1}

\usepackage{graphicx}
\usepackage{dcolumn}
\usepackage{hyperref}

\begin{document}

\newcommand{\epl}{Europhys. Lett.}
\newcommand{\sust}{Superc. Sci. Tech.}

\title{Nodeless two-gap superconductivity in stoichiometric iron pnictide LiFeAs }

\author{H.~Kim}
\affiliation{The Ames Laboratory, Ames, IA 50011, USA}
\affiliation{Department of Physics \& Astronomy, Iowa State University, Ames, IA 50011, USA}

\author{M.~A.~Tanatar}
\affiliation{The Ames Laboratory, Ames, IA 50011, USA}

\author{Yoo~Jang~Song}
\affiliation{Department of Physics, Sungkyunkwan University, Suwon, Gyeonggi-Do 440-746, Republic of Korea}

\author{Yong~Seung~Kwon}
\affiliation{Department of Physics, Sungkyunkwan University, Suwon, Gyeonggi-Do 440-746, Republic of Korea}

\author{R.~Prozorov}
\email[Corresponding author: ]{prozorov@ameslab.gov}
\affiliation{The Ames Laboratory, Ames, IA 50011, USA}
\affiliation{Department of Physics \& Astronomy, Iowa State University, Ames, IA 50011, USA}

\date{26 September 2010}

\begin{abstract}
The variations of in- and inter- plane London penetration depths, $\Delta\lambda(T)$, were measured using a tunnel diode resonator in single crystals of the intrinsic pnictide superconductor LiFeAs. This compound appears to be in the clean limit with a residual resistivity of 4 ($T\to0$) to 8 ($T_c$) $\mu \Omega\cdot$cm and $RRR$ of 65 to 35, respectively. The superfluid density, $\rho_s(T)=\lambda^2(0)/\lambda^2(T)$, is well described by the self-consistent two-gap $\gamma-$model. Together with the previous data, our results support the universal evolution of the superconducting gap from nodeless to nodal upon departure from optimal doping. We also conclude that pairbreaking scattering plays an important role in the deviation of the low-temperature behavior of $\lambda(T)$ from exponential in Fe-based compounds.
\end{abstract}



\maketitle


Studies of the superconducting gap structure play an important role in the determination of the mechanism responsible for superconducting pairing. In iron-based superconductors \cite{Kamihara2008}, the situation regarding the gap structure remains controversial. In the "1111" RFeAs(O,F) compounds (R = rare earth) at optimal doping  tunneling \cite{Chen2008} and angle-resolved photoemission spectroscopy (ARPES) \cite{Kondo2008} experiments have found a full superconducting gap. An exponential low-temperature variation of the penetration depth, $\lambda(T)$, was reported in compounds with magnetic rare earths Sm1111 \cite{Malone2009} and Pr1111 \cite{Hashimoto2009a}, but the non-magnetic La1111 has shown a close to $T^2$ power-law behavior, incompatible with a clean, full gap \cite{Martin2009}. A similar power-law behavior of $\lambda(T)$ was found in optimally  electron -doped Ba(Fe$_{1-x}$T$_x$)$_2$As$_2$ ("BaT122", T = transition metal) \cite{Gordon2009a,Williams2009,Martin2010a,Kim2010b} and hole-doped Ba$_{1-x}$K$_x$Fe$_2$As$_2$ ("BaK122") \cite{Hashimoto2009,Martin2009a}. It was suggested that such a non-exponential behavior comes from a full gap with pair-breaking scattering \cite{Dolgov2009,Bang2009,Mishra2009a,Vorontsov2009,Kogan2009,Hashimoto2009,Gordon2010,Kim2010b}. On the other hand, in clean (as suggested by observations of quantum oscillations \cite{Carrington2009,Shishido2010,Terashima2010}) isoelectronic P-doped BaFe$_2$(As$_{1-x}$P$_x$)$_2$ ("BaP122") \cite{Hashimoto2010a}, and low-$T_c$ stoichiometric LaFePO \cite{Fletcher2009} and KFe$_2$As$_2$ ("K122") \cite{Dong2010,Hashimoto2010}, the gap appears to be nodal, which was suggested to be an intrinsic behavior of clean Fe-pnictides. However, nodal behavior was observed also in dirty systems. Doping-dependent gap anisotropy \cite{Luo2009,Tanatar2010} and nodes \cite{Martin2010a,Reid2010} were reported for 122 crystals. It was suggested that a full gap at optimal doping evolves into a 3D nodal structure when doped toward the edges of the superconducting``dome''. Hereinafter, when we refer to a ``full gap'', we cannot exclude the possibility of some angular variation (e.g., see Refs.~[\onlinecite{Zeng2010,Chubukov2010,Vorontsov2010}]), and only mean that such variation is smaller than the gap magnitude.

Since doping inevitably introduces scattering \cite{Kemper2009}, which is pairbreaking in iron pnictides \cite{Dolgov2009,Bang2009,Vorontsov2009,Kogan2009,Gordon2010,Kim2010b}, measurements of stoichiometric intrinsic superconductors become of utmost importance.
LiFeAs with high $T_c \approx 18$~K \cite{Wang2008,Tapp2008,Chu2009,Song2010} is among very few such compounds. It is one of the cleanest systems with a high residual resistivity ratio ($RRR$) of about 50 \cite{Song2010}, much higher than BaP122 (5 to 8 for different doping) \cite{Kasahara2010}, BaCo122 (3 to 4) \cite{Ni2008} and BaK122 (7 to 10) \cite{Luo2008}, pure Ba122 (7 to 10 under pressure) \cite{Colombier2009}, though still below pure K122 (over 1000) \cite{Terashima2010}. Since $T_c$ of LiFeAs decreases with pressure \cite{Chu2009,Gooch2009}, which is observed only in optimally and overdoped compounds \cite{Colombier2010}, we can assign its "equivalent" doping level as slightly overdoped, as opposed to underdoped NaFeAs, whose $T_c$ goes through a maximum with pressure \cite{Zhang2009} and heavily overdoped K122. This doping assignment is consistent with the temperature-dependent resistivity, discussed later. With the much reduced effect of pairbreaking scattering, comparison of these stoichiometric compounds can bring an insight into the intrinsic evolution of the superconducting gap.

High chemical reactivity leading to quick degradation in air makes LiFeAs challenging to study. A single, fully isotropic gap \cite{Inosov2010}, as well as two-gap $s_{\pm}$ pairing \cite{Borisenko2010} were reported from ARPES. Two-gap superconductivity is also supported by magnetization \cite{Sasmal2010,Song2010a}, specific heat \cite{Wei2010} and nuclear magnetic resonance (NMR) \cite{Li2010}.

In this Letter, we report on the in-plane London penetration depth, $\lambda(T)$, in single crystals of LiFeAs measured using a tunnel diode resonator (TDR) \cite{Degrift1975}. The superfluid density can be well fitted with the self-consistent clean two-gap $\gamma-$model \cite{Kogan2009a}. Our results imply that the ground state of pnictide superconductors in the clean limit at optimal doping is given by $s_{\pm}$ symmetry with two distinct gaps, $\Delta_1/T_c \sim 2$ and $\Delta_2/T_c \sim 1$.

\begin{table*}[tb]
\caption{\label{tab1}Sample properties and parameters of exponential and power-law fits. (See text)}
\begin{ruledtabular}
\begin{tabular}{c|ccccccc}
sample & $T_c^\textmd{\scriptsize onset} (K)$&$T_c$ (K)& $\Delta T_c$ (K)& $n$&  $A'$ (pm/K$^{n}$) &$A$ (pm/K$^{3.1}$)  &$\Delta_0$ ($k_BT_c$) \\ \hline
\#1 & $18.2\pm 0.1$ & $17.5 \pm 0.1$ & $0.7 \pm 0.2$ & $3.39\pm0.04$& $64.8\pm4.8$& $106.1\pm0.7$&$1.09\pm0.02$\\
\#2 & $18.1\pm 0.1$& $17.2\pm0.1$ & $0.9\pm0.2$ &$2.96\pm0.04$&$136.9\pm10.2$ & $107.1\pm 0.7$  &$0.94\pm0.02$\\
\#3& $18.0\pm0.1$& $16.9\pm0.1$ & $1.1\pm0.2$ &$3.05\pm0.06$&$119.7\pm12.4$ & $110.1\pm0.9$& $0.95\pm0.02$

\end{tabular}
\end{ruledtabular}
\end{table*}

Single crystals of LiFeAs were grown in a sealed tungsten crucible using Bridgeman method \cite{Song2010,Song2010a} and were transported in sealed ampoules. Immediately after opening, $(0.5-1) \times (0.5-1) \times (0.1-0.3)$ mm$^3$ pieces of the same crystal (all surfaces cleaved in Apiezon N grease) were used for TDR, transport and magnetization measurements. Small resistance contacts ($\sim 0.1 m\Omega$) were tin-soldered \cite{Tanatar2010a} and resistivity was measured using a four probe technique in {\it Quantum Design} PPMS. The transition temperature, $T_c$, was determined at the maximum of the derivative $d\Delta\lambda (T)/dT$, Table~\ref{tab1}. The London penetration depth was measured with the TDR technique (for review, see \cite{Prozorov2006}). The sample was inserted into a 2 mm inner diameter copper coil that produced an rf excitation field (at $f \approx 14$~MHz) with amplitude $H_{ac} \sim 20$ mOe, much smaller than $H_{c1}$. Measurements of the in-plane penetration depth, $\Delta \lambda_{ab}(T)$, were done with $H_{ac} \parallel c$-axis, while with $H_{ac} \bot c$ we measured $\Delta \lambda_{c,mix}(T)$ that contains a linear combination of $\lambda_{ab}$ and $\lambda_{c}$ \cite{Martin2010a}. The shift of the resonant frequency, $\Delta f(T)=-G4\pi\chi(T)$, where $\chi(T)$ is the differential magnetic susceptibility, $G=f_0V_s/2V_c(1-N)$ is a constant, $N$ is the demagnetization factor, $V_s$ is the sample volume and $V_c$ is the coil volume. The constant $G$ was determined from the full frequency change by physically pulling the sample out of the coil. With the characteristic sample size, $R$, $4\pi\chi=(\lambda/R)\tanh (R/\lambda)-1$, from which $\Delta \lambda$ can be obtained \cite{Prozorov2000,Prozorov2006}.

\begin{figure}[tb]
\includegraphics[width=1.0\linewidth]{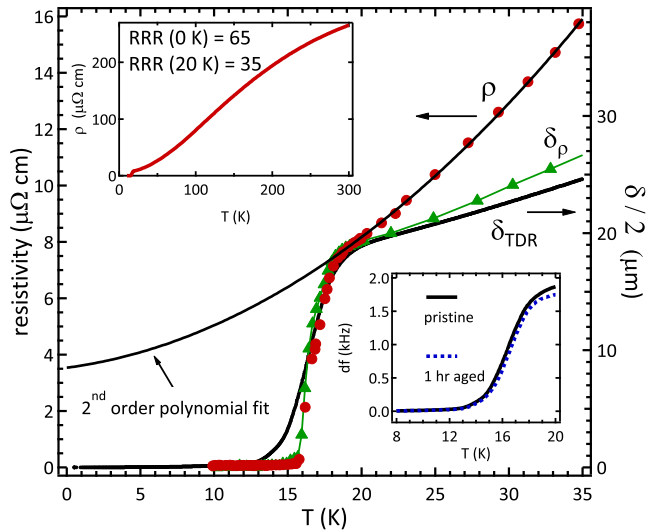}%
\caption{\label{fig1}Left axis: resistivity (symbols) along with the 2$^{nd}$ order polynomial fit from $T_c$ to 50 K used to determine the residual resistivity, $\rho (0) = 3.7~\mu \Omega \cdot$cm. Right axis: skin depth, $\delta$, measured by TDR, $\delta_{TDR}$ compared to that calculated from resistivity, $\delta_{\rho}$. Upper inset: $\rho(T)$ in the full temperature range. Lower inset: TDR data for pristine and air-aged samples (see text).}
 \end{figure}

The main panel in Fig.~\ref{fig1} shows the temperature-dependent resistivity, $\rho(T)$ (left axis), and skin depth, $\delta (T)$ (right axis). $\rho(T)$ up to room temperature is shown in the top inset. The residual resistivity ratio, $RRR = \rho (300K)/\rho (20~K) = 35$ and it reaches the value of 65 when extrapolated to $T=0$ using a  2$^{nd}$ order polynomial. This behavior is consistent with the $T$-dependent resistivity of BaCo122 in the overdoped regime \cite{Doiron-Leyraud2009}. The calculated skin depth, $\delta _{\rho} (T) = (c/2\pi) \sqrt{\rho/f}$, compares well with the TDR data for $T > T_c$, where $\Delta f/f_0 = G [1-\Re \{ \tanh{(\alpha R)/(\alpha R)}\}]$, $\alpha = (1-i)/\delta$ \cite{Hardy1993} when we use $\rho(300K)$=250 $\mu \Omega\cdot$cm, the lowest directly measured value among our crystals. A very good quantitative match of two independent measurements gives us a confidence in both resistivity data and the TDR calibration.

To check for degradation effects, a sample was intentionally exposed to air for an hour and the measurements were repeated, as shown in the lower inset in Fig.~\ref{fig1}. After the exposure, the sample surface lost its shiny metallic gloss, but the transition temperature and width remained the same. At the same time, the total frequency shift through the transition (proportional to the sample volume) decreased, suggesting that the surface had degraded without introducing any change in the bulk. This experiment clearly supports bulk uniform superconductivity of our samples.

\begin{figure}[tb]
\includegraphics[width=1.0\linewidth]{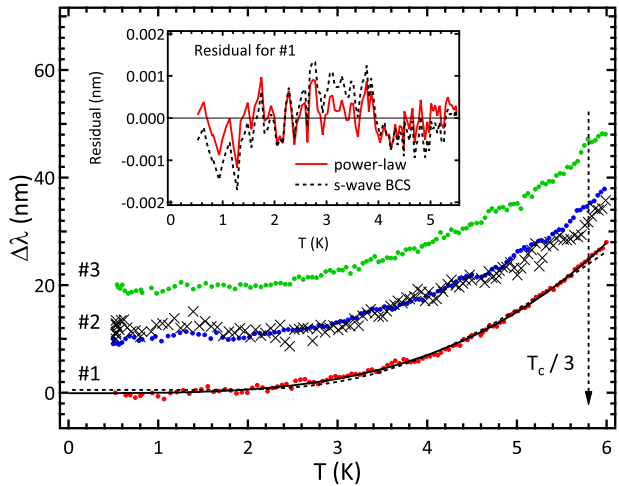}%
\caption{\label{fig2}Main panel: $\Delta \lambda_{ab} (T)$ in three LiFeAs crystals (solid dots) and $\Delta \lambda_{c,mix} (T)$ for sample \#2 (crosses). Analysis (shown for \#1) was done assuming both power-law (solid lines) and exponential (dashed line) $T-$ dependences. The data for samples \#2 and \#3 (shifted vertically for clarity by 10 and 20 nm, respectively) were analyzed in a similar way, see Table~\ref{tab1}. Inset: comparison of the fit residuals for sample \#1 for the power-law and exponential functions.}
 \end{figure}

$\Delta \lambda_{ab} (T)$ in three LiFeAs crystals is shown up to $T_c/3$ in Fig.~\ref{fig2} by solid dots. $\Delta \lambda_{ab} (T)$ was analyzed using (1)  power-law, $\Delta \lambda (T)=AT^n$ (with $A$ and $n$ being free parameters) and (2) exponential BCS form, $\Delta\lambda(T)=\tilde{\lambda_0}\sqrt{\pi\Delta_0/2T}\exp{(-\Delta_0/T)}$ (with $\tilde{\lambda_0}$ and $\Delta_0$ as free parameters). The best fit results for sample \#1 are shown with solid (power-law) and dashed (exponential) lines. The fit residuals are shown in the inset. The exponential fit quality is as good as the power-law, although $\Delta_0/T_c = 1.09 \pm 0.02$ is smaller than the value of 1.76 expected for a conventional single fully-gapped s-wave pairing and $\tilde{\lambda_0} = 280 \pm 15$ nm is somewhat larger than the experimental $200$ nm \cite{Pratt2009,Inosov2010}. This is naturally explained by two-gap superconductivity in LiFeAs. The superconducting $T_c$ and best fit parameters (obtained from fitting up to $T_c/3$) for all samples are summarized in Table~\ref{tab1}. $T_c^\textmd{\scriptsize onset}$ was defined at 90\% of the rf susceptibility variation over the transition: the mean $T_c$ was defined at the maximum of $d\Delta\lambda(T)/dT$ and $\Delta T_c = T_c^\textmd{\scriptsize onset} - T_c$. $\Delta_0/T_c$ from the single-gap exponential BCS behavior. The power-law coefficient $A'$ was obtained with the exponent $n$ as a free parameter, while $A$ was obtained with a fixed $n=3.1$ (average of 3 samples). Crosses in Fig.~\ref{fig2} show $\Delta \lambda_{c,mix} (T)$ for sample \#2. A clear saturation of $\Delta \lambda_{c,mix} (T)$ at low temperatures suggests exponential behavior of $\lambda_{c}$.

 \begin{figure}[tb]
\includegraphics[width=1.0\linewidth]{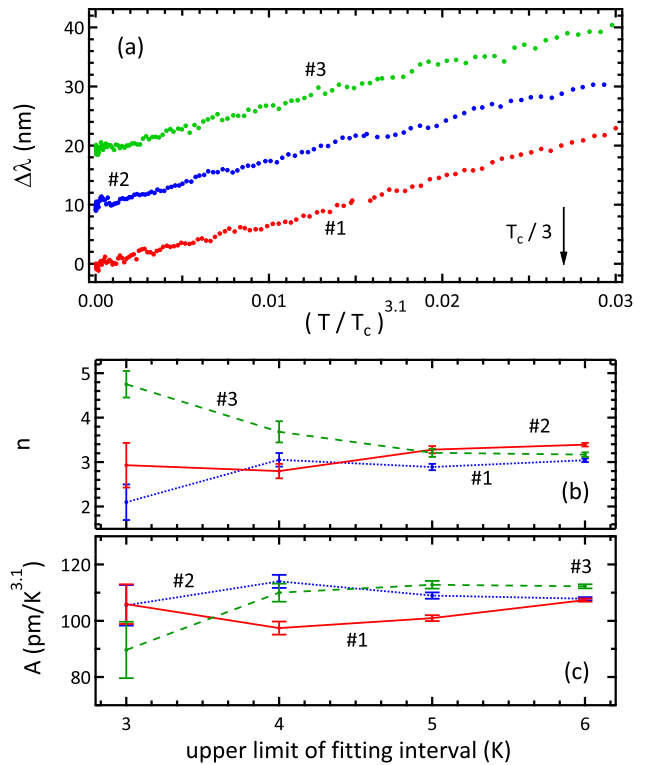}%
\caption{\label{fig3}(a) $\Delta \lambda$ vs. $T^{3.1}$ (average exponent $n$ over 3 samples); (b) exponent $n$ and (c) prefactor $A$, obtained by fitting to $\Delta\lambda(T)\propto AT^n$ for various upper temperature limits shown on the $x$-axis.}
 \end{figure}

Figure~\ref{fig3}(a) shows $\Delta \lambda(T)$ vs. $T^n$  with $n=3.1$ which is the average exponent for three samples. The dependence of the parameters, $n$ and $A$, on the fitting temperature range is summarized in Fig.~\ref{fig3} (b) and (c), respectively. As expected, the exponent $n$ is more scattered for the shortest fit interval, otherwise $n$ and $A$ do not depend much on the fitting range from base temperature to 6 K and give $n>3$ for all samples, with the average value $3.13\pm0.23$. With $n$ fixed at this average value, we determined the prefactor $A = 107.8\pm2.1$ $\mu$m/K$^{3.1}$.

\begin{figure}[tb]
\includegraphics[width=1.0\linewidth]{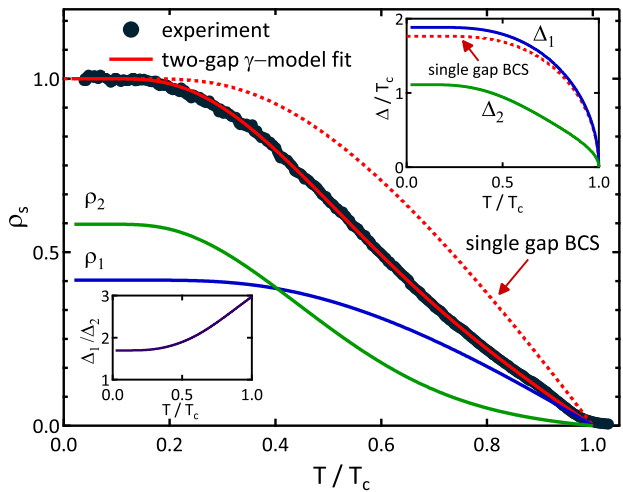}%
\caption{\label{fig4}Symbols: superfluid density, $\rho_s(T)$ calculated with $\lambda(0)=200$ nm. Solid lines represent the fit to a two-gap $\gamma-$model, $\rho_s = \gamma \rho_1 + (1-\gamma) \rho_2$. Dashed line is a single-gap BCS solution. Upper inset: superconducting gaps, $\Delta_1(T)$ and $\Delta_2(T)$ calculated self-consistently during the fitting. Lower inset: $\Delta_1/\Delta_2$ as a function of temperature.}
\end{figure}

The superfluid density, $\rho_s (T)=[1 + \Delta \lambda(T)/\lambda(0)]^{-2}$ is the quantity to compare with the calculations for different gap structures. Figure~\ref{fig4} shows $\rho_s (T)$ for crystal \#1 calculated with $\lambda(0)=200$ nm \cite{Pratt2009,Inosov2010}. A noticeable positive curvature above $T_c/2$ is similar to other Fe-based superconductors \cite{Prozorov2009} and MgB$_2$ \cite{Fletcher2005}. This suggests multigap superconductivity in LiFeAs, which we analyze in the framework of the self-consistent $\gamma-$model \cite{Kogan2009a}. Since LiFeAs is a compensated metal and its bandstructure is fairly two-dimensional \cite{Borisenko2010}, we may introduce the mass ratio on two cylindrical bands, $\mu = m_1/m_2$, whence the partial density of states of the first band, $n_1=\mu/(1+\mu)$. The total superfluid density is $\rho_s = \gamma \rho_1 + (1-\gamma) \rho_2$ with  $\gamma=1/(1+\mu)$. We also use the Debye temperature of 240 K \cite{Wei2010} to calculate $T_c$, which allows fixing one of the interaction constants, $\lambda_{11}$. This leaves three free fit parameters: the in-band, $\lambda_{22}$, and interband, $\lambda_{12}$, interaction and the mass ratio, $\mu$. Figure~\ref{fig4} shows that $\rho_s(T)$ can be well described in the entire temperature range by this clean-limit weak-coupling BCS model. In the fitting, the two gaps were calculated self-consistently (which is the major difference between this one and the popular, but not self-consistent, $\alpha$ - model \cite{Bouquet2001}) and the self-consistent $\Delta_1(T)$ and $\Delta_2(T)$ are shown in the upper inset in Fig.~\ref{fig4}, while the gap ratio is shown in the lower inset indicating strong non-single-gap-BCS behavior of the small gap. The best fit, using Levenberg-Marquardt nonlinear simplex  optimization in \emph{Matlab}, gives $\Delta_1(0)/T_c \sim 1.885$ and $\Delta_2(0)/T_c \sim 1.111$. As expected, one of the gaps is larger and the other is smaller than the single-gap value of 1.76, which is \emph{always} the case for a self-consistent two-gap solution. The best fit parameters are: $\lambda_{11} = 0.630$, $\lambda_{22} = 0.642$, $\lambda_{12} = 0.061$ and $\mu = 1.384$. The determined mass ratio gives $n_1 = 0.581$ and $\gamma = 0.419$. This is consistent with bandstructure calculations that yield $n_1 = 0.57$ and $\mu = 1.34$ \cite{Mazin2010a}, and ARPES experiments that find $\mu \approx 1.7$ \cite{Borisenko2010}. The effective coupling strength, $\lambda_{eff} = 0.374$, is not far from $0.35$ estimated for 122 \cite{Boeri2010} and $0.21$ for 1111 \cite{Boeri2008} pnictides. The electron band with a smaller gap gives about 1.5 times larger contribution to the total $\rho_s$ resulting in a crossing of the partial densities at low temperatures. Similar result was obtained from magnetization measurements \cite{Sasmal2010}. We stress that while $\Delta_1(T)$, $\Delta_2(T)$ and $\mu$ (hence, $n_1$, $n_2$ and $\gamma$) and $\lambda_{eff}$ are unique self-consistent solutions describing the data, the coupling matrix $\lambda_{ij}$ is not unique. There are other combinations that could produce similar results, so it seems that $\lambda_{ij}$ has to be calculated from first principles \cite{Mazin2010a}.



In conclusion, we find that in the clean limit, optimally-doped Fe-based superconductors are fully gapped, but most measurements are affected by pairbreaking scattering \cite{Dolgov2009,Bang2009,Mishra2009a,Vorontsov2009,Kogan2009,Hashimoto2009,Gordon2010,Kim2010b}. This conclusion is in line with studies of thermal conductivity (which is not so sensitive to scattering) in BaCo122 \cite{Tanatar2010,Reid2010}. On the other hand, intrinsic K122 reveals a nodal gap \cite{Dong2010,Hashimoto2010}, which is also found in heavily overdoped BaCo122 \cite{Reid2010}. Overall, this establishes a common trend for all Fe-based superconductors to have a superconducting gap that evolves from full to nodal when moving toward the edge of the superconducting dome. 


We thank S. Borisenko, S. Bud'ko, P. Canfield, A. Chubukov, D. Evtushinsky, P. Hirschfeld, V. Kogan, Y. Matsuda and I. Mazin for useful discussions. Work at Ames was supported by the U.S. Department of Energy, Office of Basic Energy Science, Division of Materials Sciences and Engineering. Ames Laboratory is operated for the U.S. Department of Energy by Iowa State University under Contract No. DE-AC02-07CH11358. Work at SKKU was partially supported by Basic Science Research Program through the National Research Foundation of Korea(NRF) funded by the Ministry of Education, Science and Technology (2010-0007487). R.P. acknowledges support from the Alfred P. Sloan Foundation.

\end{document}